\documentclass[aps,prl,superscriptaddress,twocolumn,twoside,a4paper,floatfix]{revtex4-1}
\usepackage{hyperref,graphicx,amsmath,amssymb,mathpazo,color}
\usepackage[normalem]{ulem}

\usepackage{pgfplots,tikz}

\usepackage{amsfonts,epsfig}
\usepackage{hyperref}
\usepackage{latexsym}
\usepackage{epsfig}
\usepackage{amsmath}
\usepackage{amssymb}
\usepackage{amsfonts}
\usepackage{amsthm}
\usepackage{mathrsfs}
\usepackage{natbib}
\usepackage{color,verbatim,graphics}
\usepackage{psfrag}
\DeclareMathAlphabet{\mathrsfs}{U}{rsfs}{m}{n}
\DeclareMathAlphabet{\mathpzc}{OT1}{pzc}{m}{it}
\DeclareMathAlphabet{\matheus}{U}{eus}{m}{n}
\DeclareMathAlphabet{\mathbbold}{U}{bbold}{m}{n}

\newcommand{\ba}{\begin{eqnarray}}
\newcommand{\be}{\begin{equation}}
\newcommand{\ee}{\end{equation}}

\newcommand{\ea}{\end{eqnarray}}
\newcommand{\ban}{\begin{eqnarray*}}
\newcommand{\ean}{\end{eqnarray*}}
\newcommand{\Tr}{\operatorname{Tr}}

\newcommand{\ket}[1]{|#1\rangle}

\newcommand{\ketbra}[2]{|#1\rangle\langle#2|}

\begin{document}

\title{Certifying the dimension of classical and quantum systems in a prepare-and-measure scenario with independent devices}

\author{Joseph Bowles}
\affiliation{D\'epartement de Physique Th\'eorique, Universit\'e de Gen\`eve, 1211 Gen\`eve, Switzerland}
\author{Marco T\'ulio Quintino}
\affiliation{D\'epartement de Physique Th\'eorique, Universit\'e de Gen\`eve, 1211 Gen\`eve, Switzerland}
\author{Nicolas Brunner}
\affiliation{D\'epartement de Physique Th\'eorique, Universit\'e de Gen\`eve, 1211 Gen\`eve, Switzerland}
\affiliation{H.H. Wills Physics Laboratory, University of Bristol, Bristol, BS8 1TL, United Kingdom}

\begin{abstract}
We consider the problem of testing the dimension of uncharacterised classical and quantum systems in a prepare-and-measure setup. Here we assume the preparation and measurement devices to be independent, thereby making the problem non-convex. We present a simple method for generating nonlinear dimension witnesses for systems of arbitrary dimension. The simplest of our witnesses is highly robust to technical imperfections, and can certify the use of qubits in the presence of arbitrary noise and arbitrarily low detection efficiency. Finally, we show that this witness can be used to certify the presence of randomness, suggesting applications in quantum information processing. 
\end{abstract}

\maketitle

The problem of estimating the dimension of uncharacterised physical systems has recently attracted attention. From a fundamental point of view, this problem is well motivated, as it shows that dimension---the number of (relevant) degrees of freedom---of an unknown system can be determined in a device-independent way. That is, dimension can be tested from measurement data alone, in a scenario in which all devices used in the experiment, including the measurement device, are uncharacterised, i.e. no assumption about the internal working of the devices is needed. Beyond the fundamental interest, this problem is also relevant in the context of quantum information, where the dimension of quantum systems---i.e. the Hilbert space dimension---represents a resource for performing information-theoretic tasks. Specifically, higher dimensional quantum systems can increase the performance of certain protocols, and/or simplify their implementation.

First approaches to this problem considered Bell inequality tests \cite{hilbert,tamas,david,YC,jens,Miguel}, random access codes \cite{wehner}, and monitoring of an observable of a dynamic system \cite{wolf}. More recently, a general formalism was developed to estimate the dimension of classical and quantum systems in a prepare-and-measure setup \cite{gallego}, the simplest but also the most general scenario. Consider two uncharacterised devices, hence described as black boxes (see Fig. 1). The first device prepares upon request a physical system in an unknown state $\rho_x$. A second device then performs a measurement on the system. The observer tests the devices, by choosing a preparation $x$ and a measurement $y$, then receiving measurement outcome $b$. Repeating the experiment many times, the observer obtains the probability distribution $p(b|x,y)$, called here the data. The goal for the observer is then to give a lower-bound on the dimension of the unknown set of states $\{\rho_x\}$ from the data alone. This can be achieved using ``dimension witnesses'' \cite{gallego,BNV,arno} (see also \cite{stark, harrigan} for different approaches). These ideas were shown to be relevant experimentally \cite{hendrych,ahrens}, and for quantum information processing \cite{marcin,li}.

 \begin{figure}
 \includegraphics[width=0.8\columnwidth]{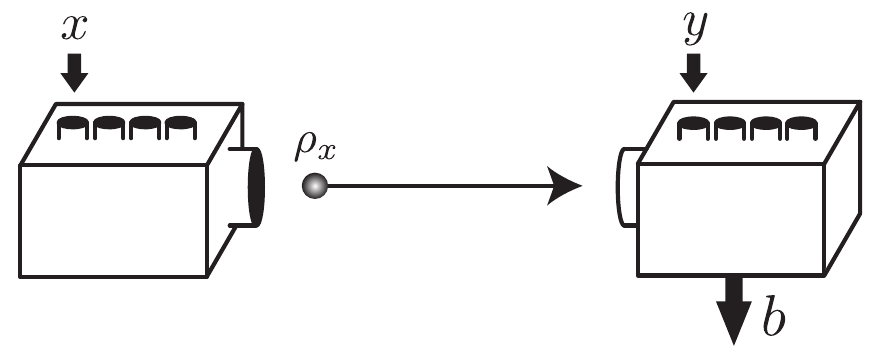}
  \caption{Prepare-and-measure setup.}
\label{fig}
\end{figure}

Here we discuss this problem assuming the preparation and measurement devices to be independent. This assumption is rather natural in a device-independent estimation scenario, where devices are uncharacterised but do not conspire maliciously against the observer. The main difficulty of this problem is that it is non-convex, a feature that makes generic problems with independent variables hard to tackle. Note that previous works on dimension witnesses allowed the devices to be correlated via shared randomness (hence relaxing the independence assumption), making the problem convex. Although these techniques can in principle be applied in our case, they are far from optimal, as we shall see below. 

It is therefore desirable to develop novel methods, which is the goal of this work. Specifically, we present a simple technique for deriving nonlinear dimension witnesses, tailored for device-independent tests of dimension assuming independent devices. We derive witnesses for systems of arbitrary dimension, obtaining a quadratic gap between classical and quantum dimensions. The simplest witness is discussed in detail. We show that it is extremely robust to technical imperfections, and can be used to certify the presence of randomness.

\emph{Scenario.} We consider the setup of Fig.1. The experiment is characterised by the set of conditional probabilities $p(b|x,y)$ (i.e. the data) which gives the probability of obtaining outcome $b$ when performing measurement $y$ on preparation $x$. 

Consider first the case of quantum systems. We say that the experiment admits a $d$-dimensional quantum representation when there exist states $\rho_x$ and measurement operators $M_{b|y}$ both acting on $\mathbb{C}^d$, such that 
\ba \label{Q} p(b|x,y) = \Tr(\rho_x M_{b|y}). \ea
Next consider the situation of classical systems of dimension $d$. Given the choice of preparation $x$, the first device sends a classical message $m=0,...,d-1$. Note that the device may have an internal source of randomness (represented by a random variable $\lambda_1$). Hence, which message $m$ is sent depends on both $x$ and $\lambda_1$. 
The measurement device, upon receiving message $m$, and input $y$ from the observer, delivers an outcome $b$. As it also features a source of randomness 
(random variable $\lambda_2$), the output $b$ depends on $m$, $y$, and $\lambda_2$. 
The behaviour observed in the experiment is then given by
\small
\ba \label{SR} p(b|x,y) = \int d \lambda_1 d \lambda_2 \rho(\lambda_1,\lambda_2) \sum_{m=0}^{d-1} p(m|x,\lambda_1) p(b|m,y,\lambda_2). \nonumber \ea
\normalsize
The main point now is to consider the joint distribution of random variables $\lambda_{1,2}$. If $\rho(\lambda_1,\lambda_2) \neq \rho_1(\lambda_1) \rho_2(\lambda_2)$, the variables are correlated, hence the devices may follow a (pre-established) correlated strategy. Previous works focused on this situation. As the set of behaviours of the above form is convex, it can be fully characterised with linear dimension witnesses \cite{gallego}.

Here we consider the situation in which the devices are independent, i.e. $\rho(\lambda_1,\lambda_2) = \rho_1(\lambda_1) \rho_2(\lambda_2)$. That is, although each device features an internal source of randomness, the devices have no shared randomness. In this case, the observed statistics can be written as
\ba \label{NSR} p(b|x,y) = \sum_{m=0}^{d-1} s(m|x) t(b|m,y) \ea
where $s(m|x)= \int d \lambda_1 \rho_1(\lambda_1) p(m|x,\lambda_1)$ is the distribution of possible messages $m$ for each preparation $x$, and $t(b|m,y)= \int d \lambda_2 \rho_2(\lambda_2) p(b|m,y,\lambda_2)$ is the distribution of outcomes $b$ for measurement $y$ when receiving message $m$. Below we will see how to characterise the set of behaviours of the form \eqref{NSR}. This will require nonlinear dimension witnesses as the set is non-convex.

\emph{Determinant witness.} In this work we focus on experiments with binary outcomes, denoted $b=0,1$. We will construct nonlinear witnesses based on the determinant of a matrix. We first discuss the simplest case, with four preparations $x=0,...,3$ and two measurements $y=0,1$. Consider the following matrix 
\begin{equation}\label{wbit}
\mathbf{W}_2=
\begin{pmatrix}
p(0,0)-p(1,0) && p(2,0)-p(3,0) \\
p(0,1)-p(1,1) && p(2,1)-p(3,1)
\end{pmatrix}
\end{equation}
where we write $p(x,y) = p(b=0|x,y)$ for simplicity. For any strategy involving a classical bit (i.e. its statistics admits a decomposition of the form \eqref{NSR} with $d=2$), one has that
\ba W_2=\text{det}( \mathbf{W}_2)=0. \ea
The proof is straightforward. Note that for any statistics of the form \eqref{NSR} with $d=2$, we have that $p(x,y)=s(0|x)[t(0|0,y)-t(0|1,y)]+t(0|1,y)$. Hence we write
\small
\ba   p(x,y)-p(x',y)&=&[s(0|x)-s(0|x')][t(0|0,y)-t(0|1,y)]  \nonumber \\ &=& S_{xx'}T_y  \ea
\normalsize
from which it follows that
\begin{equation}\label{wbit}
W_2=
\begin{vmatrix}
S_{01}T_0 && S_{23}T_0 \\
S_{01}T_1 && S_{23}T_1
\end{vmatrix}=0.
\end{equation}
An interesting feature of the above witness is that it is given by an equality, whereas linear witnesses are given by inequalities \cite{gallego}. Moreover, our witness turns out to characterise fully the set of experiments involving a classical bit. Specifically, for any experiment achieving $W_2=0$ (for all relabelings of the preparation $x$), there exists a decomposition of the form \eqref{NSR} with $d=2$ (see Appendix A). Note that if the preparation and measurement devices are correlated, then classical bit strategies can reach $W_2=1$. Consider for instance the equal mixture of the two following deterministic strategies: (i) $s(0|x)=1$ iff $x=0,3$ and $t(0|m,y)=m+y$ mod $2$, (ii) $s(0|x)=1$ iff $x=0,2$ and $t(0|m,y)=m$. Hence we get $\mathbf{W}_2= \mathbb{I}_2$ and $W_2=1$. This shows that our witness is tailored for the case in which the devices are independent.

Next we investigate the performance of qubit strategies, i.e. statistics of the form \eqref{Q} with $d=2$. States are given by density matrices $\rho_x=(\mathbb{I}_2+\vec{s}_x\cdot\vec{\sigma})/2$ and measurement operators by $M_{0|y}=c_{y}\mathbb{I}_2+\vec{T}_y\cdot\vec{\sigma}/2$, where $\vec{s}_x$ and $\vec{T}_y$ are Bloch vectors and $|c_y|\leq1$ \cite{mparam}. Similarly to above, we write 
\ba p(x,y)-p(x',y)=\text{Tr}[(\rho_{x}-\rho_{x'})M_{0|y}]=\vec{S}_{xx'}\cdot \vec{T}_{y}\;\; \ea
where $\vec{S}_{xx'}=(\vec{s}_{x}-\vec{s}_{x'})/2$. Finally, we get
\begin{equation}\label{wbit}
W_2= 
\begin{vmatrix}
\vec{S}_{01}\cdot \vec{T}_{0} && \vec{S}_{23}\cdot \vec{T}_{0} \\
\vec{S}_{01}\cdot \vec{T}_{1} && \vec{S}_{23}\cdot \vec{T}_{1}
\end{vmatrix}= (\vec{S}_{01}\times\vec{S}_{23})\cdot(\vec{T}_{0}\times\vec{T}_{1}) \leq 1
\end{equation}
since $|\vec{S}_{01}\times\vec{S}_{23}|\leq 1$ and $|\vec{T}_{0}\times\vec{T}_{1}|\leq 1$. This bound for qubit strategies is tight, and can be reached as follows: choose the preparations to be the pure qubit states given by $\vec{s}_0=-\vec{s}_1=\hat{z}$, $\vec{s}_2=-\vec{s}_3=\hat{x}$, and the measurements by the vectors $\vec{T}_0=\cos{\theta}\hat{z}+ \sin{\theta}\hat{x}$ and $\vec{T}_1=\sin{\theta}\hat{z}- \cos{\theta}\hat{x}$. Notice that we are free to choose any angle $\theta$ here, due to the rotational invariance of the cross product in the plane. For $\theta=0$ we get the usual BB84 states and measurements.

It is relevant to note that essentially any qubit strategy achieves $|W_2| > 0$. Only very specific alignments of the qubit preparations and measurements (a set of measure zero) achieve $W_2=0$. Therefore, a generic qubit strategy always outperforms the most general strategy involving a bit.

This suggests that our witness is well-suited for distinguishing data involving classical bits and qubits. To illustrate the robustness of our witness, we investigate the effect of technical imperfections, such as background noise and limited detection efficiency (of the detector inside the measurement device), on a generic qubit strategy given by the data $p_Q(x,y)$ achieving $|W_2|=Q > 0$. Say that an error occurs with probability $1-\eta$, for instance the emitted particle is lost. Hence the observed statistics is given by 
\ba p(x,y)= \eta p_Q(x,y)  + (1-\eta) p_N(y), \ea
where we consider a noise model of the form $p_N(x,y)=p_N(y)$, i.e. the noise is independent of the choice of preparation $x$. The difference in probabilities entering the witness is then independent of the noise term: $p(x,y)-p(x',y)= \eta [p_Q(x,y)-p_Q(x',y) ]$, and thus the observed value of the witness is $W_2=\eta^2 Q$, which is strictly positive whenever $Q >0$. Hence, for an arbitrary amount of background noise and/or an arbitrarily low efficiency, a generic qubit strategy will outperform any classical bit strategy; see \cite{attacks} for a related result. This is indeed in stark contrast with previous witnesses, which can only tolerate a finite amount of noise and require a high efficiency \cite{arno}.

Finally, we comment on strategies involving higher dimensional systems. Using a classical trit one achieves $|W_2| \leq 1$ \cite{trit}, while numerical analysis shows that $|W_2| \leq 1.299 $ for qutrit strategies. This shows that the value of $W_2$ is useful to assess dimension. To reach the algebraic maximum of $W_2=2$, systems of dimension (at least) $d=4$ (either classical or quantum) are required.

\emph{Determinant witness for all dimensions.} We now generalise the above witness for testing classical and quantum systems of arbitrary dimension. Consider a scenario with $2k$ preparations and $k$ binary measurements. Construct the $k \times k$ matrix 
\ba \mathbf{W}_{k}(i,j)=p(2j,i)-p(2j+1,i) \ea 
with $0\leq i,j \leq k-1 $. As above, the witness is given by $W_{k}=|\text{det}(\mathbf{W}_k)|$. We will see that, for classical systems of dimension $d$, one has that
\ba W_k=0 \quad \text{for   } d\leq k, \ea
while one can have $W_k\geq 1$ for $d>k$. For quantum systems of dimension $d$, we get 
\ba W_k=0 \quad \text{for   } d\leq \sqrt{k} ,\ea
while $W_k>0$ is possible whenever $d>\sqrt{k}$. Hence we obtain a quadratic separation between classical and quantum dimensions, using a number of preparations and measurements that grows only linearly.

To prove the above claims, it is enough to focus on quantum strategies. Consider matrices of the form
\ba \label{gm}
\rho_x=\frac{1}{d}\bigg(\mathbb{I}_{d}+\phi_d\;\vec{s}_x \cdot\vec{\lambda}\bigg),
\ea
with $\vec{s}_x\in\mathbb{R}^{d^2-1}, |\vec{s}_x|\leq1$, $\vec{\lambda}$ the vector of the $d^2-1$ Gell-Mann matrices (generalised Pauli matrices, satisfying $\text{tr}(\lambda_i)=0$ and $\text{tr}(\lambda_i\lambda_j)=2\delta_{ij}$) and  $\phi_d=\sqrt{(d(d-1))/2}$. 
While all matrices of the above form are valid quantum density matrices for $|\vec{s}_x|\leq2/d$ \cite{kimura}, this is not the case in general (although this will not affect our argument).
Similarly we write measurement operators as $M_{0|y}= c_y \mathbb{I}_{d}+\phi_d\;\vec{T}_{y}\cdot\vec{\lambda}/d$ with $\vec{T}_y\in\mathbb{R}^{d^2-1}, |\vec{T}_y|, |c_y|\leq1$ \cite{mparam}, and get that 
\begin{align} \label{prod}
\mathbf{W}_{k}(i,j)=\text{Tr}[(\rho_{2j}-\rho_{2j+1})M_{0|i}]=\vec{S}_{j}\cdot\vec{T}_{i}
\end{align}
with $\vec{S}_{j}=(1-\frac{1}{d})(\vec{s}_{2j}-\vec{s}_{2j+1})$. Thus, as before, the entries of the matrix $\mathbf{W}_{k}$ are given by scalar products of vectors. 
Similarly to the qubit construction of eq. \eqref{wbit}, the witness $W_k$ can be expressed using cross products, generalised here to arbitrary dimensions. 

Specifically, the cross product $\vec{S}_{0}\times\vec{S}_{1}\times\cdots\times\vec{S}_{k-1}$ of $k$ vectors in $\mathbb{R}^{k+1}$ is defined as the unique vector $\vec{u}\in \mathbb{R}^{k+1}$ such that $\vec{V}\cdot\vec{u}=\text{det}(\vec{S}_{0},\vec{S}_{1},\cdots,\vec{S}_{k-1})$ for all $\vec{V}\in\mathbb{R}^{k+1}$ (see e.g. \cite{dittmer}). It follows that $\vec{S}_{0}\times\cdots\times\vec{S}_{k-1}=0$ iff $\vec{S}_0, \cdots, \vec{S}_{k-1}$ are linearly dependent. Furthermore, similarly to eq. \eqref{wbit}, we have that
\small
\begin{align}\label{gencross} \nonumber
W_k = |\text{det}(\mathbf{W}_k)| = |(\vec{S}_{0}\times\cdots\times\vec{S}_{k-1})\cdot(\vec{T}_{0}\times\cdots\times\vec{T}_{k-1})|
\end{align}
\normalsize
To conclude, we relate the dimension of the quantum systems to the linear (in)dependence of the set of vectors $\vec{S}_{j}$ and $\vec{T}_{i}$. 
Note that we must ensure here that the vectors $\vec{S}_{j}$, $\vec{T}_{i}$ are in $\mathbb{R}^{k+1}$, via an embedding or by using only a restricted set of parameters.
As $d$-dimensional quantum systems have $d^2-1$ parameters, we see that the vectors $\vec{S}_{j}$ (and similarly for $\vec{T}_{i}$) can span a subspace of dimension at most $d^2-1$. Hence, if $d \leq \sqrt{k}$, the vectors $\vec{S}_{j}$ cannot be linearly independent, and we get $W_k=0$. On the contrary if $d> \sqrt{k}$, the vectors $\vec{S}_{j}$ and $\vec{T}_{i}$ can be chosen to be linearly independent, and we have $W_k>0$. Take for instance $\vec{S}_{j}$ to be parallel to $\vec{T}_{i}$, and $|\vec{s}_j|,|\vec{T}_i|\leq2/d$ ensuring that all preparations and measurements are represented by valid operators. Note however that this construction is suboptimal in general, as one can obtain $W_k=1$ with quantum states of dimension $d>\sqrt{k}$ (with $d$ an integer prime power), using mutually unbiased basis (see Appendix B). 

The proof for classical systems can be derived by noting that any classical strategy using $d$-dimensional states can be recast as a quantum strategy using diagonal density matrices acting on $\mathbb{C}^d$. Since we have only $d-1$ parameters in this case, it follows from the above that $W_k=0$ when $d\leq k$. For $d> k$, one can get $W_k\geq 1$. The lower bound is obtained by considering the following strategy: if $x$ is even, then send $m=x/2$, else send $m=d$; for the measurement device, output $b=0$ iff $y=m$. Note that for this strategy, we get $\mathbf{W}_{k}=\mathbb{I}_{k}$, hence $W_k=1$. An interesting question is to find the algebraic maximum of $W_k$, and the minimal dimension for classical and quantum systems required to attain it. Note that this problem is related to that of finding the determinant of a Hadamard matrix. Hence we get the bound $W_k \leq k^{k/2}$, which is tight iff there exists a Hadamard matrix of size $k\times k$.

\emph{Certifying randomness.} The fact that the determinant witness can distinguish between classical and quantum systems (given a bound on the dimension) suggests applications in randomness certification. Here we investigate the connection between the amount of violation of the witness $W_2$ and the intrinsic randomness of the of the underlying statistics, assuming that the preparation device emits qubit states.

Consider the quantity 
\ba
\bar{p} = \frac{1}{4} \sum_{x,y=0,1} \max_{b} p(b|x,y), \ea 
i.e. the average guessing probability of the outcome $b$ for preparations $x=0,1$. Randomness can be quantified by the min-entropy of $\bar{p}$, i.e. $H_{min}(\bar{p})= -\log_2(\bar{p})$, which gives the number of random bits extractable from the experiment (per run). Now for a given amount of violation of the witness $W_2=Q>0$, we want to find out the maximal value of $\bar{p}$ over all qubit strategies which are compatible with the value $W_2=Q>0$. In other words, what is the minimal amount of randomness compatible with a certain violation of the witness? To answer this question, we solve numerically the following problem. We maximise $\bar{p}$ subject to the constraints: $W_2=Q$, $p(b|x,y) = \Tr(\rho_x M_{b|y})$ where $ \rho_{x}$, $M_{b|y}$ are arbitrary qubit states and measurement operators. 

In Figure 2, we plot the amount of randomness $H_{min}(\bar{p})$ as a function of the value $Q$ of the witness $W_2$. We see that for any amount of violation, randomness can be certified. In other words, from the sole knowledge of the value of $W_2$, one can upper bound the probability of correctly guessing the output $b$, for any observer knowing the detailed qubit strategy that is being used. Importantly, the quantity $H_{min}(\bar{p})$ captures here the intrinsic quantum randomness of the experiment, but is independent of any randomness generated locally in the devices (used e.g. to create mixed state preparations). These issues will be discussed in detail in a forthcoming work \cite{bowles}, where a protocol for randomness certification will be presented.

\begin{figure}
    \centering
    \includegraphics[scale=0.95]{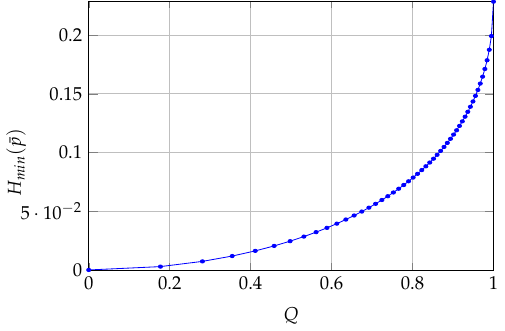} 
    \caption{Average certifiable randomness $H_{min}(\bar{p})$ using the witness $W_{2}$. For any amount of violation of the witness $W_2=Q>0$, randomness can be certified.}
    \label{random}
\end{figure}

\emph{Discussion.} We have presented a method for testing the dimension of classical and quantum systems of arbitrary dimension. Moreover, the simplest of our witnesses is highly robust to noise and can be used to certify randomness without the need of high visibilities and efficiencies. Hence we believe these ideas are relevant in practice. In this perspective, it will be necessary to make a statistical analysis in the spirit of Refs \cite{gill} for taking finite size effects into account \cite{bowles}. Finally, from a more abstract point of view, the ideas presented here could be useful in other non-convex problems involving independent variables, such as Bell tests with independent sources \cite{bilocality,fritz}, and more general marginal problems \cite{pearl}.

\emph{Acknowledgements.} We thank C. Ci Wen Lim, M. Paw\l owski, S. Pironio, E. Woodhead, and H. Zbinden for discussions, and acknowledge financial support from the Swiss National Science Foundation (grant PP00P2\_138917) and the EU DIQIP.

\section{Appendix}

\textit{A. Full characterisation of the set of classical bit strategies with the witness $W_2$}--- In the main text we showed that $W_{2}=0$ for strategies involving a classical bit. Here we will see that the converse holds. That is, any statistics achieving $W_2=0$ for all possible relabelings of the preparation label $x$, can be realised with a classical bit strategy. 

Consider the matrix $\mathbf{W}_{2}$, here rewritten as 
\begin{equation}\label{iffmatrix}\mathbf{W}_{2}=
\begin{pmatrix}
a_1-a_2 && b_1 - b_2  \\
c_1 - c_2  && d_1 -d_2
\end{pmatrix}
\end{equation}
with $a_1=p(0,0)$, $a_{2}=p(1,0)$ and so on. Without loss of generality, we take $d_2\geq d_1\geq c_2\geq c_1$ which can be achieved via a relabelling of the preparations $x$. The conditions $W_{2}=0$, considering all relabelings of $x$, are then given by 
\begin{align} \label{cond} (a_1-a_2)(d_1-d_2)&=&(b_1-b_2)(c_1-c_2) \nonumber \\
(a_1-b_2)(d_1-c_2)&=&(b_1-a_2)(c_1-d_2) \\
(a_1-b_1)(c_2-d_2)&=&(a_2-b_2)(c_1-d_1) \nonumber \end{align} 
where the second and third equations correspond to relabelling $x$ according to $1\leftrightarrow 3$ and $1 \leftrightarrow 2$ respectively. To show that there exists a decomposition of the form (2) with $d=2$, we solve for
$\{s(m|x)\}, \{t(0|m,y\}$ (with $m=0,1$, $x=0,...,3$, and $y=0,1$) the set of equations given by the conditions \eqref{cond} and the 8 conditions given by $p(x,y) = \sum_{m=0,1} s(m|x) t(0|m,y)$. A solution is given by $t(0|0,0)=b_1$, $t(0|0,1)=d_1$, $t(0|1,0)=b_2$, $t(0|1,1)=d_2$, $s(0|0)=(d_2-c_1)/(d_2-d_1)$, $s(0|1)=(d_2-c_2)/(d_2-d_1)$, $s(0|2)=1$ and $s(0|3)=0$. Hence any matrix $(\ref{iffmatrix})$ satisfying conditions \eqref{cond} admits a decomposition of the form \eqref{NSR} with $d=2$. Thus the determinant witness characterises fully the set of distributions obtained from strategies involving a classical bit. 

\textit{B. Quantum strategy using mutually unbiased bases}--- Consider a quantum system of dimension $d$, for which we have $n\leq d+1$ mutually unbiased bases (MUBs) denoted by ${\cal{M}}_{\alpha}=\{\ket{\psi_{i | \alpha}}\}$, where $\alpha=0,\cdots,n-1$ and $i=0,\cdots,d-1$. Due to the properties of MUBs the projectors $\pi_{i | \alpha}=\ketbra{\psi_{i | \alpha}}{\psi_{i | \alpha}}$ satisfy $\text{tr}(\pi_{i | \alpha}\pi_{j | \alpha})=\delta_{ij}$ and $\text{tr}(\pi_{i | \alpha}\pi_{j | \beta})=1/d$ for $\alpha\neq\beta$. The main idea now will be to construct a quantum strategy for which we get $\mathbf{W}_{k}=\mathbb{I}_{k}$ and so $W_{k}=1$. 

Consider first the upper left block of $\mathbf{W}_{k}$ of size $d-1\times d-1$. Concentrating on the first basis ${\cal{M}}_{0}$, we choose the preparations as $\rho_{2j}=\pi_{j|0}$ and $\rho_{2j+1}=\pi_{d-1|0}$, (with $j=0,\cdots,d-2$) and measurement projectors as $M_{0|i}=\pi_{i|0}$, (where $i=0,\cdots,d-2$). Hence for this block we have that
\begin{equation}
p(2j,i)-p(2j+1,i)=\text{tr} ([\pi_{j|0}-\pi_{d-1|0}]\pi_{i|0})=\delta_{ij}
\end{equation}
since ${\cal{M}}_{0}$ is an orthonormal basis, and so the first $d-1\times d-1$ block of $\mathbf{W}_{k}$ is the identity matrix $\mathbb{I}_{d-1}$. 

We then move on to the next $d-1$ preparations and measurements, keeping the same pattern but using the next basis ${\cal{M}}_{1}$. That is, we choose $\rho_{2j+2(d-1)}=\pi_{j|1}$, $\rho_{2j+1+2(d-1)}=\pi_{d-1|1}$ and $M_{0|i+d-1}=\pi_{i|1}$  $(i,j=0,\cdots,d-2)$. We continue this pattern until we have used up all $n$ MUBs. This will give us a $(d-1)n\times (d-1)n$ matrix. Via the same argument as above, each $d-1\times d-1$ block on the diagonal of $\mathbf{W}_{k}$ will be equal to $\mathbb{I}_{d-1}$. All off-diagonal blocks are the zero matrix since they contain preparations and measurements that belong to different MUBs. Indeed we have that $\text{tr}( [\pi_{i|\alpha}-\pi_{j|\alpha}]\pi_{k|\beta})=0$ when $\alpha \neq \beta$.

Hence when $(d-1)n\geq k$, we get that $\mathbf{W}_{k}=\mathbb{I}_{k}$, hence $W_k = 1$. If the Hilbert space dimension $d$ is an integer power of a prime, then there exist $d+1$ MUBs \cite{WootersMUB}. In this case, one has that $W_k = 1$ for $d\geq\sqrt{k+1}$, or equivalently $d>\sqrt{k}$.


\begin{thebibliography}{99}

\bibitem{hilbert} N. Brunner, S. Pironio, A. Ac\'\i n, N. Gisin, A.A. M\'ethot, V. Scarani, Phys. Rev. Lett. {\bf 100}, 210503 (2008).

\bibitem{tamas} K.F. P\'al, T. V\'ertesi, Phys. Rev. A {\bf 77}, 042105 (2008).

\bibitem{david} D. P\'erez-Garc\'ia, M.M. Wolf, C. Palazuelos, I. Villanueva, M. Junge, Comm. Math. Phys. {\bf 279}, 455 (2008).

\bibitem{YC} T. Moroder, J.-D. Bancal, Y.-C. Liang, M. Hofmann, and O. G\"uhne, Phys. Rev. Lett. {\bf 111}, 030501 (2013).

\bibitem{jens} C. Eltschka and J. Siewert, Phys. Rev. Lett. {\bf 111}, 100503 (2013).

\bibitem{Miguel} M. Navascu\'es, G. de la Torre, T. V\'ertesi, Phys. Rev. X {\bf 4}, 011011 (2014).

\bibitem{wehner} S. Wehner, M. Christandl, A.C. Doherty, Phys. Rev. A {\bf 78}, 062112 (2008).

\bibitem{wolf} M.M. Wolf, D. P\'erez-Garc\'ia, Phys. Rev. Lett. {\bf 102}, 190504 (2009).

\bibitem{gallego} R. Gallego, N.  Brunner, C. Hadley, A. Ac\'\i n, Phys. Rev. Lett. \textbf{105}, 230501 (2010).



\bibitem{BNV} N. Brunner, M. Navascu\'es, T. V\'ertesi, Phys. Rev. Lett. {\bf 110}, 150501 (2013).

\bibitem{arno} M. Dall'Arno, E. Passaro, R. Gallego, A. Acin, Phys. Rev. A {\bf 86}, 042312 (2012).

\bibitem{stark} C. Stark, arXiv:1209.5737.

\bibitem{harrigan} N. Harrigan, T. Rudolph, S. Aaronson, arXiv:0709.1149.

\bibitem{hendrych} M. Hendrych, R. Gallego, M. Micuda, N. Brunner, A. Ac\'\i n, J. Torres, Nat. Phys. {\bf 8}, 588 (2012).

\bibitem{ahrens} J. Ahrens, P. Badziag, A. Cabello, M. Bourennane, Nat. Phys. {\bf 8}, 592 (2012).


\bibitem{marcin} M. Paw\l owski, N. Brunner, Phys. Rev. A {\bf 84}, 010302(R) (2011).

\bibitem{li} H.-W. Li \emph{et al.}, Phys. Rev. A {\bf 85}, 052308 (2012); Phys. Rev. A {\bf 84}, 034301, (2011).

\bibitem{attacks} M. Dall'Arno, E. Passaro, R. Gallego, M. Pawlowski, A. Acin, arXiv:1210.1272.

\bibitem{kimura} G. Kimura, Phys. Lett. A {\bf 314}, 339 (2003)

\bibitem{dittmer} A. Dittmer, American Math. Monthly {\bf 101}, 887 (1994).

\bibitem{bowles} J. Bowles \emph{et al.}, in preparation.


\bibitem{gill} R. Gill, arXiv:1207.5103; Y. Zhang, S. Glancy, and E. Knill, arXiv:1303.7464.


\bibitem{bilocality} C. Branciard, N. Gisin, and S. Pironio, Phys. Rev. Lett. {\bf 104}, 170401 (2010).

\bibitem{fritz} T. Fritz, New J. Phys. {\bf 14}, 103001 (2012).

\bibitem{pearl} J. Pearl, \emph{Models, reasoning and inference}, Cambridge Univ. Press (2000).

\bibitem{WootersMUB} W. Wooters, B. Fields, Annals of Phys. {\bf 191}, 363 (1989).

\bibitem{mparam} Note that not all matrices with $|c_{y}|\leq 1$ are valid measurement operators since $c_{y}$ is in general dependant $\vec{T}_{y}$.

\bibitem{trit} Although qubits and classical trits perform equally for the witness $W_{2}$, qubits are able to outperform trits for $W_{3}$ (defined later).
\end{thebibliography}
\end{document}